\documentstyle[prl,aps,epsf,multicol]{revtex}
\input{psfig}
\begin{document}

\title{Phase Structure of Systems with Multiplicative Noise}

\author{G. Grinstein, M.A. Mu{\~n}oz, and Yuhai Tu}

\address{IBM Research Division, T.J.
Watson Research Center, P.O. Box 218, Yorktown Heights, NY 10598}
\date{\today}
\maketitle

\begin{abstract}
The phase diagrams and
transitions of nonequilibrium systems with
multiplicative noise are studied theoretically.
We show the existence of both strong and weak-coupling
critical behavior, of two distinct active phases, and
of a nonzero range of
parameter values over which the susceptibility is infinite
in any dimension.  A scaling theory of the
strong-coupling transition is constructed.
\end{abstract}

\pacs{64.60.Ht, 02.50.-r, 47.20.Ky}
\begin{multicols}{2}
\narrowtext

Though they have been argued\cite{Brand} to
describe a diverse and important set of physical systems out of
equilibrium,
among them autocatalytic chemical processes and several different
problems in quantum optics,
dissipative partial differential equations (pde's)
with multiplicative noise remain poorly understood.
Such equations typically admit two types of phases:
trivial or "absorbing" phases in which the dynamical variable(s)
or order parameter(s) vanish identically at all points in space,
and remain zero in perpetuity; and nontrivial or "active" phases
in which these variables have nonzero expectation values.
To date, most of the theoretical effort on systems
of this type
has been devoted\cite{Brand,Kramer} to calculating the critical
value of the control parameter at which the continuous transition between
these two phases occurs\cite{Graham}.

In this paper we present a somewhat broader
investigation of the multiplicative noise problem,
along the lines of existing analyses of critical phenomena in the
directed percolation problem\cite{DP},
and in the many physically-relevant
stochastic
pde's with additive noise.
In particular, we analyze the
phase structure and critical properties
of multiplicative-noise problems with
different symmetries.
We show that above an upper critical dimension,
$d_c =2$, the transition can belong to one of two distinct universality
classes described by different fixed points: a weak coupling, mean-field
fixed point,
accessible for noise strength $D$ less than a (nonuniversal) critical
value $D_c$; and a strong coupling
fixed point with nontrivial exponents, accessible for $D>D_c$.
The separatrix at $D=D_c$ is described by a third, unstable
fixed point, likewise with nontrivial exponents.
For $d \leq 2$, only the strong coupling
critical behavior occurs.  We analyze the mean-field transition,
compute the critical exponents on the separatrix in an expansion
in $\epsilon \equiv d-2$, and
formulate a scaling description of the
strong coupling transition,
which we check against previous exact results
for $d=0$ (the single-variable problem).
We also demonstrate the occurrence
of a rather striking phenomenon: For discrete symmetry and
any $d$,
there is a region
of the phase diagram in which the response of the system
to an infinitesimal uniform field (i.e., the uniform susceptibility),
is everywhere
infinite.  For systems with N-component
order parameters and $O(N)$ symmetry with $N \geq 2$,
we argue for the existence of two distinct active phases for $d > 2$,
one which breaks
the $O(N)$ symmetry and one which does not.  We illustrate this
with an exact calculation for $N= \infty$.

The models we study are defined by the equation

\begin{equation}
\partial_t n_{i} (\vec x ,t) = \mu \nabla^2 n_{i}
- r n_{i}
- u|\vec n|^2 {n_{i}}^{\rho} /N + |\vec n|  \eta_{i} ~~.
\label{defn}
\end{equation}
Here $\vec n \equiv \{n_1 , ..., n_N \}$
is an N-component, real vector field, $r$ and $u~ (>0)$ are
real parameters, and $\vec \eta$ is a Gaussian noise vector
with correlations
$<\eta_i (\vec x ,t) \eta_j
(\vec x' ,t') > = D  \delta_{i j} \delta (\vec x -\vec x')
\delta (t-t') /N$, for some noise strength $D$.  We deal primarily
with the cubic nonlinearity
$\rho =1$, though
for the single-component theory, $N=1$, we
also discuss
aspects of the quadratic nonlinearity,
$\rho=0$.  We work exclusively in the Ito\cite{VK} representation.

To establish some of the general phenomenology
of these models, we first discuss mean-field
theory.  The neglect of fluctuations which is the
essence of mean field is most simply implemented
by dropping the noise term in (\ref{defn}), and
ignoring spatial variations in $\vec n(\vec x ,t)$, which
is replaced by $\vec n(t)$.  For $N=1$,
Eq. (\ref{defn}) then has
two steady-state solutions with $dn/dt =0$, respectively
representing the absorbing and active phases:
$n(t) =0$, stable for $r>0$, and $n = (-r/u)^{1/(1+\rho)}$,
stable for $r<0$.  For the continuous symmetry case, $N \geq 2$,
(where only the cubic nonlinearity, $\rho =1$, preserves the
O(N) symmetry and so need be considered),
a result identical to that for $N=1$, $\rho =1$
holds, except that in the active phase the orientation of $\vec n$
is unspecified, the magnitude being given by ${\vec n}^2 = (-r/u)$.
In both cases the transition between the phases occurs at $r=r_c=0$,
and the critical exponent $\beta$ characterizing the vanishing
of $n$ as $r \rightarrow r_c$
takes the value $\beta = 1/ (1+\rho)$, independent of
dimension, $d$.  The correlation length and dynamical exponents
assume their usual mean-field values, $\nu =1/2$ and $z=2$,
respectively.

Alternatively, one can formulate mean-field theory in terms of a
model with infinite-range interactions (or equivalently, $d=\infty$),
by replacing
$\mu \nabla^2 n$ in (\ref{defn})
by $\alpha (<\vec n> -\vec n)$, where $\alpha$
is a positive constant and
$<\vec n>$ the expectation value of $\vec n$.
For $N=1$, this model can be solved self-consistently for
$<\vec n>$, giving a continuous transition at $r_c =0$
with an exponent $\beta = 1/(1+\rho)$.

It is particularly easy to determine the critical dimension
$d_c$ above which this mean-field result is valid, by considering
the absorbing
phase where $<\vec n(\vec x ,t)> = 0$.
In this phase, one can calculate
certain correlation and
response functions exactly, by summing perturbation
expansions to all orders\cite{Peliti}.  We first specialize to
$N=1$, where
the exact renormalized noise, $D_R$, and coupling constant $u_R$,
are given by
$D_R = D A$, and $u_R = u A $,
with $A=1/(1-DI_d (r))$, and $I_d (r) \equiv
\int \frac{d^d k}{(2\pi )^d} \frac{1}{2(\mu k^2 +r)}$.
Note that $I_d(r)$ is finite
for all $r \geq 0$, provided $d > 2$\cite{UV}.  Hence
for small enough $D$ the denominator of $A$ remains bounded
below by a positive number, even when $r$ decreases to zero.
Thus $D_R$ and $u_R$ remain finite and nonzero throughout
the absorbing phase, right down to the critical point, which
continues to occur precisely at $r=0$, provided $D$ is small enough
and $d>2$.  In this regime, all other quantities of interest
likewise experience no singular renormalization near $r=0$, so the
mean-field
critical point at $r=0$ and mean-field exponents remain exact.
(In fact some quantities, such as the response function $g(\vec k ,
\omega ) = -i\omega + \mu k^2 + r$, experience no
diagrammatic renormalization whatsoever
in the absorbing phase\cite{Peliti}.)

For $d \leq 2 $, by contrast, $I_d (r)$ diverges at $r=0$,
so $D_R$ and $u_R$ both
blow up for some {\it positive} value of $r$ even for
infinitesimal $D$.
This invalidates
mean-field theory, and suggests that the critical
value $r_c$ is shifted away from zero, and that the mean-field critical
exponents can no longer be trusted.  The same is true for $d>2$,
provided $D$ exceeds the special value $D_c$ for which
$D_R$ first diverges at $r=0$.  These results strongly suggest that
$d_c=2$, but that even for $d>d_c$ there may be nontrivial critical
behavior at large $D$ \cite{Pik}.

This expectation is confirmed by straightforward
renormalization-group (RG) analysis\cite{WK}
of models (\ref{defn}), in the absorbing phase and at the critical point.
Imagine rescaling space, time, and the field according to
$\vec x = b \vec x '$, $t=b^z t'$, and $\vec n (\vec x ,t) = b^{\zeta}
\vec n' (\vec x',t')$, where $z$ and $\zeta$ are as yet undetermined
exponents.  Writing $b=e^l$, and using standard methods,
and the results quoted
above, one readily derives the following recursion relations

\begin{eqnarray}
&d&\mu /dl = (z-2) \mu \nonumber \\
&d&r/dl = 2r \nonumber \\
&d&u/dl = u((1+\rho) \zeta + z + (1+2\rho) A_d D ) + O(\rho u^2)
\nonumber \\
&d&D/dl  = D(z-2-\epsilon + A_d D ) ~~.
\label{rr}
\end{eqnarray}

Here $\epsilon = d-2$, and
$A_d =1/4\pi +O(\epsilon)$ is a positive, d-dependent constant.
Owing to
the aforementioned absence of any diagrammatic renormalization
of the response function in the absorbing phase,
the $\mu$ and $r$ equations
are exact.
The $u$ and $D$
equations contain no further diagrammatic corrections
in the case
$\rho=0$.  For $\rho =1$, there are, as indicated in (\ref{rr}),
corrections to the
$u$ equation of $O(u^2)$.

Obviously, finite fixed points of these
recursion relations can only be reached by choosing $z =2$ and
tuning $r(l=0)$,
the initial value
of $r$, to zero.  Choosing
$\zeta =
-(z+(1+2\rho)A_d D )/(1+\rho)$ maintains the $u$ equation at
a fixed point with $u^\ast = u(l=0)$.
Then one need only look for stable
fixed points of the $D$ equation.

For $d>2$ ($\epsilon >0$),
this equation has a stable ``weak-coupling"
fixed point at $D^\ast = 0$, which supplements the other
fixed point values: $r^\ast =0$,
and $u^\ast =u(l=0)$.
This fixed point gives rise to mean-field exponents
(e.g., $z=2$ and
$\nu =1/2$ follow immediately from the nonrenormalization of
the response function),
and is reached
for initial values of $D$ less than $D_c = \epsilon / A_d $.

For $D > D_c$, however,
the recursion relation for $D$ runs off to $D=\infty$, rendering
the critical behavior incalculable by perturbative techniques.
Presumably the
transition is controlled by a nonperturbative ``strong-coupling"
fixed point in this regime.  For $d \leq 2$, however, the weak-coupling
fixed point with $D^\ast =0$ is unstable for {\it any} positive $D$,
and strong-coupling
behavior always obtains.

While the strong-coupling transition cannot yet be
described in full detail,
certain aspects of it have been elucidated in previous work.
In particular, Becker and Kramer\cite{Kramer} have
derived results for the amount, $r_c$, by which the critical value of $r$
is shifted away from $r=0$ at this transition.  They find that $r_c <0$
(in the Ito representation) for d=1 and d=2.
There is also a rather complete solution\cite{Brand,Graham}
of the single-variable
($d=0$) problem for $N=1$, where $r_c$ is also shown to be negative.
These results allow one
to make the striking prediction that for $N=1$ the
susceptibility of model (\ref{defn}) at zero frequency ($\omega =0$)
and wavevector ($\vec k = 0$)
diverges over some nonzero range of values of $r$.
To understand this, first recall that, as discussed above,
the response function $g(\vec k ,t)$ in the absorbing phase
doesn't undergo any
diagrammatic corrections due to the nonlinearity, and so is given
by the linear result $g(\vec k ,t) = \theta (t) e^{-(\mu k^2 + r)
 t} $.  It follows at once that the susceptibility at $\vec k =
\omega = 0$, defined as $\chi =
\int_0^{\infty} dt g(\vec k = \vec 0 ,t)$, is infinite
for all negative values of $r$ that lie
in the absorbing phase.  The previous results that $r_c < 0$ for
$d \leq 2$ (which we extend to $d>2$ below),
therefore imply the divergence of $\chi$ in the entire range $0 \geq r
\geq r_c$.  For $d=0$ we demonstrate explicitly below that this range is
controlled by a fixed line with a continuously varying exponent.  Note
that the critical exponents $z$ and $\nu$ associated with the
response function
take their mean-field values, 2 and 1/2 respectively,
at the point $r=0$ where the susceptibility
diverges.  This is consistent with the recursion relations for
$r$ and $\mu$
in (\ref{rr}).  Keep in mind, however, that the strong-coupling
transition
into the active phase occurs at $r(l=0) = r_c <0$, and so is represented
by an inaccessible fixed point of the $r$ recursion relation
with $r^\ast = - \infty$.

Thus for $N=1$
we are led to the schematic phase diagrams shown in fig. 1.
In constructing the diagram
for $d>2$, we have generalized ref. \cite{Kramer}'s calculation of
$r_c$ to dimensions larger than 2.
That calculation is based on mapping the computation of $r_c$ onto the
quantum mechanical
problem of finding the lowest bound state energy of a
potential given by the spatial
correlation function of the
noise.  The analysis can easily be extended to
dimensions $d>2$, where it is well known that the
depth of a potential
well has to exceed a $d$-dependent critical value, $D_{c}$,
in order for there to be a bound state. The existence
of a bound state for $D>D_{c}$
implies that
the critical value of r is shifted to $r_c <0$, corresponding to the
strong coupling fixed point.
When $D<D_{c}$, there is no bound state, so $r_c =0$, corresponding
to the weak coupling fixed point. These results, though obtained in a
different way, are fully consistent with our earlier RG analysis, and are
summarized in fig. 1b.

\vspace{1.0in}

\psfig{file=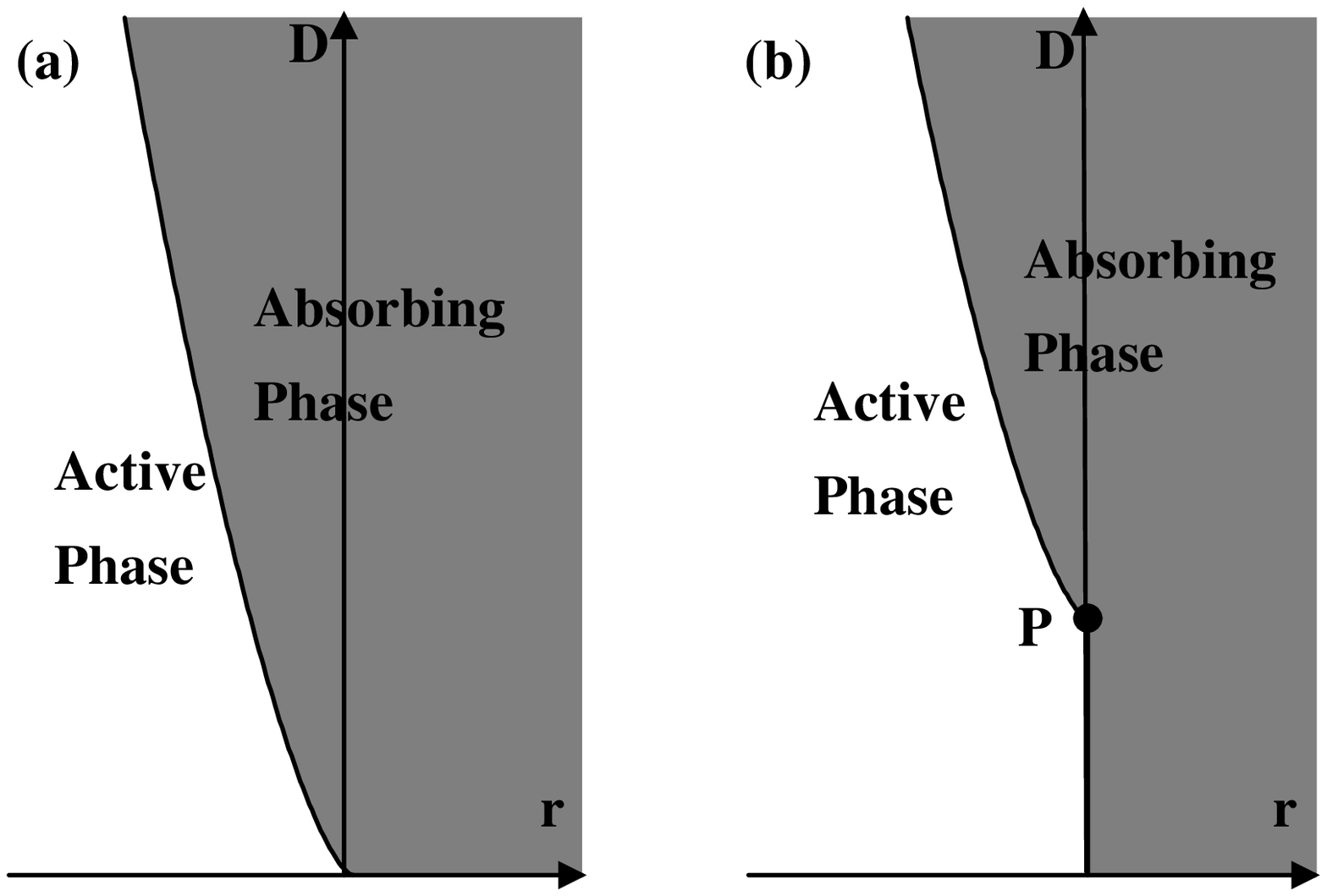,width=234pt}
\begin{figure}
\caption{Schematic phase diagram for model (1) with $N=
1$
for (a)
$d \leq 2$; (b) $d>2$.  Weak (strong) coupling transiti
on occurs
below (above) the multicritical point $P$ in (b); trans
ition is always
strong-coupling in (a).  Susceptibility diverges in the
 absorbing phases
when $r \leq 0$, and possibly also in portions of the a
ctive phases.}
\end{figure}

The multicritical point $P$ in fig. 1b is controlled by the unstable fixed
point $r^{\ast}$, $u^{\ast} = u(l=0)$,
$D^\ast = \epsilon /A_d$ of Eq. (\ref{rr}), with
$\zeta = -(z +
(1+2\rho )\epsilon )/(1+\rho )$.
The simple
scaling relation, $ M(r) = b^{\zeta} M(rb^{1/\nu}) $,
for the order parameter $M=<n>$ yields $M \sim r^{\beta}$, where $\beta
=-\zeta \nu$.  Since
$z=2$ and $\nu =1/2$, we
infer
that $\beta = (2 + (1+2\rho) \epsilon ) / 2 (1+\rho) $
for this multicritical
point.  This result is exact for $\rho = 0$,
but has corrections of
$O(\epsilon^2)$ for $\rho = 1$.

We now turn to the phase diagrams of the $O(N)$ symmetric
models with $N \geq 2$ and $\rho=1$.
The continuous symmetry of these models
allows for two distinct active phases: one which preserves
this symmetry and one which breaks it.  The first of these
has $\vec M \equiv <\vec n> =0$ and $Q \equiv
<q> \equiv <(\vec n)^2 /N > \neq 0$,
while the
second has both $\vec M \neq 0$ and $Q \neq 0$\cite{Ising}.
Since for $d \leq 2 $, it is extremely difficult to break a
continuous symmetry in a noisy system\cite{Mermin},
we anticipate that
the only active state is the symmetric one
in this case.  For $d >2$ both active phases can occur in the phase
diagram.  The absorbing phase always
occurs for sufficiently large $r$.

These features can be investigated explicitly in the exactly solvable
limit of $N =
\infty$.  In this limit, the
diagrams
contributing to the perturbation expansion for all quantities of
interest are small in number and can be easily summed, with the
following results:

\vspace{1.0in}

\psfig{file=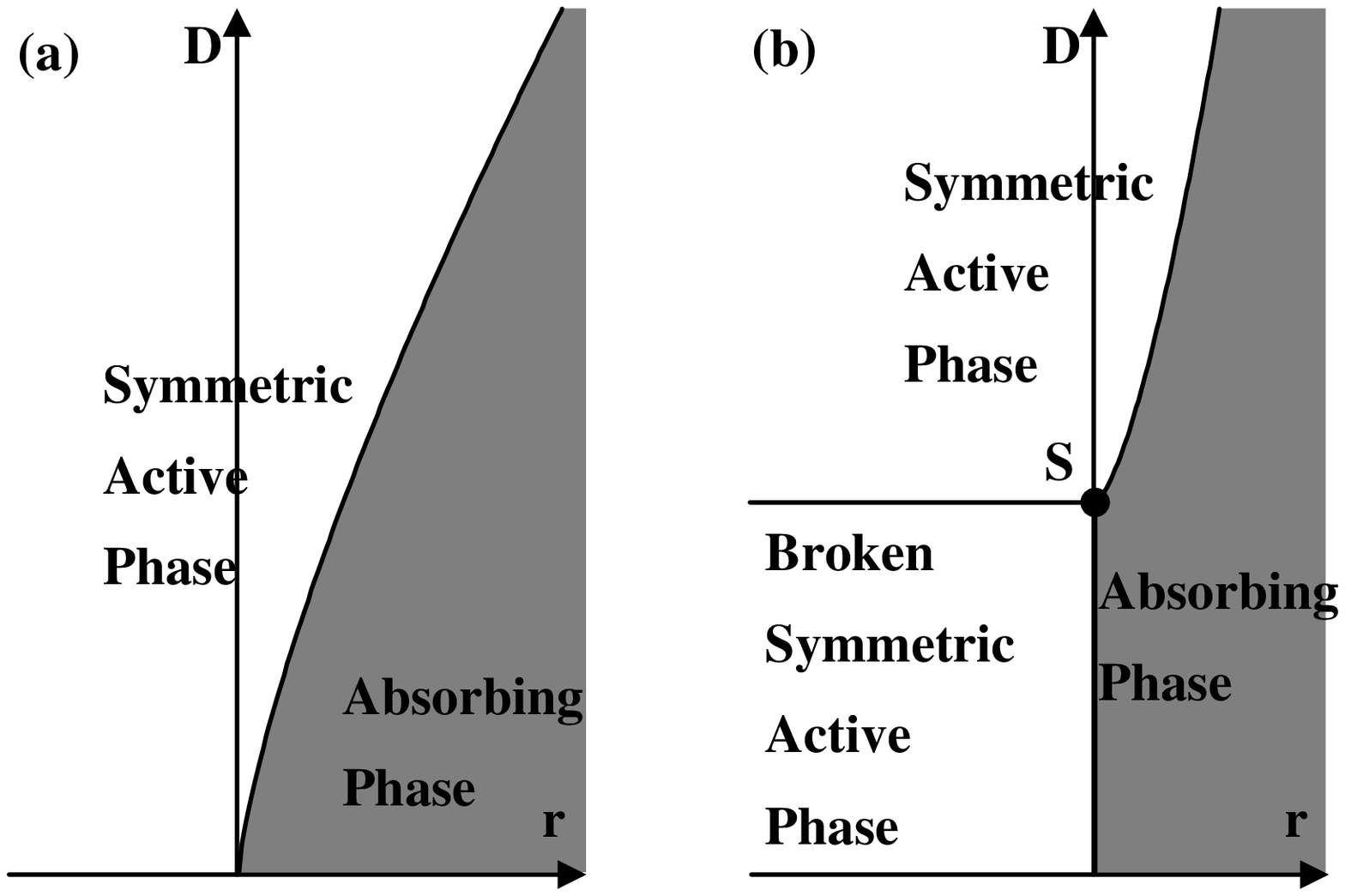,width=234pt}

\begin{figure}
\caption{Schematic phase diagram for model (1) with $N=
\infty$
for (a) $d \leq 2$; (b) $d>2$.  Symmetric (broken-symmetric)
to absorbing
transition is strong (weak) coupling in (b); transition
 is always
strong coupling in (a).}
\end{figure}

For $d \leq 2$ there is, as anticipated, no phase that breaks the
$O(N)$ symmetry, i.e.,
for which $\vec M \neq 0$.  There is, however, a symmetric active
phase with $Q > 0$.  The value of $Q$ is determined by
$D I_d (r+uQ) =1$ (where $I_d$ was defined earlier).
The phase boundary $r_c (D)$ between the absorbing and active
phases is thus
determined by $DI_d (r_c ) = 1$, so that $r_c (D) \sim D^{-2/
\epsilon}$ as $D \rightarrow 0$.  Note that $r_c$ is {\it positive} here.
Hence the susceptibility at $\vec k = \omega = 0$ remains finite
at the transition, reflecting the fact that $<\vec n > =0$
in the active phase.  Critical exponents are of course also readily
calculated.  For example,
the order parameter $Q$ vanishes like $(r_c - r) ^{\beta}$ with
$\beta = 1$.  

For $d>2$ the phase diagram is somewhat more complicated, containing
a broken-symmetric phase for small $D$, a symmetric active phase
for larger $D$ (see fig. 2b), and a multicritical point where these
active phases and the absorbing phase meet.
Again, all phase boundaries and critical
exponents can be exactly computed.  For example, the transition
from the absorbing phase to the broken-symmetry active phase is
mean-field-like, and occurs at the unshifted critical value $r_c=0$.
The exponent
governing the decay of
$Q$ at the symmetric-to-absorbing
transition continues to assume the value
$\beta = 1$, the phase boundary being determined by the equation
$DI_d (r_c) =1$, which has a solution only for $D>D_c$, with $D_c=1/I_d (0)$.
The second-order phase boundary between the two active phases
occurs at $D=D_c$ for all $r<0$, $M$ vanishing linearly
with $D_c -D$ as this boundary is approached from the broken-symmetric
phase.  The multicritical point S occurs at $r=0$, $D=D_c$ (fig. 2b).

While many details of these phase diagrams
are doubtless special to $N=\infty$,
certain qualitative features should continue to apply to
physically-relevant values of $N$ such as $N=2$.
The existence of both types of active phases for $d>2$, but only the
symmetric active phase for $d \leq 2$,
is, e.g., a general feature.  For $d>2$,
the broken-symmetric active phase will continue to occur for small $D$,
where the exponents for the transition to the absorbing phase continue
to assume mean-field values, and $r_c=0$.  The exponents
for the absorbing-to-symmetric active phase transition are presumably
controlled by a strong-coupling fixed point, and so are difficult to
calculate.

We turn next to a scaling characterization of the strong-coupling transition
for $N=1$.
This is very similar to standard scaling theories of equilibrium critical
phenomena, but since one of phases is absorbing,
an additional independent exponent is
required for a complete description of the phase transition.
To understand this, consider the
RG analysis of the steady-state two-point correlation function
$C(\vec x ,t) = < n(\vec x ,t) n(\vec 0 ,0)>$.  The rescalings of
$n,~\vec x~$ and $t$ given above yield the familiar RG equation
$C(\vec x ,t, \delta r) =
b^{2\zeta } C(\vec x /b , t/b^z ,\delta rb^{1/\nu} )$, where
$\delta r = r_c - r$.
First consider
the equal-time case, $t=0$.  Here we obtain $C(\vec x) \sim {\delta r}^{-2
\zeta \nu }
c(x/\xi)$, where $c(y)$ is a scaling function, and the correlation length
$\xi$ diverges like $\xi \sim \delta r^{-\nu}$.
In typical critical phenomena,
one can go to the critical point,
$\delta r =0$, assume that $C(\vec x)$ approaches a
nonzero value in this limit, and conclude that $C(\vec x ) \sim x^{2\zeta}$.
Here, however, because $C$ is identically zero in the absorbing
phase, it vanishes as $\delta r \rightarrow 0$ from the active phase.
Hence in the limit $\delta r \rightarrow 0$, $C(\vec x)$ takes the
form $\delta r^{\Delta} x^{-(d-2+\eta)}$, where $\Delta$ is a new,
apparently independent, critical exponent, and $\eta$ is the standard
correlation function exponent.
Consistency between this result and the previous
scaling expression requires $c(y) \sim y^{-(d-2+\eta)}$ for $y<<1$, i.e.,
$\Delta = -\nu (2\zeta + d-2+\eta )$.  This scaling law
can be written more familiarly in terms of the order parameter exponent
$\beta$, shown above to satisfy
$\beta = -\nu \zeta$, which
yields $\beta = (\Delta  +\nu (d-2+\eta))/2$.  This is a generalization
of $\beta = \nu (d-2+\eta)/2$, which holds for ordinary equilibrium critical
phenomena, where $\Delta = 0$.  In the directed percolation problem\cite{DP},
where the transition
is also into an absorbing phase, a simple graphical argument (which fails
for the current models), shows that
$\Delta = \beta$, whereupon one obtains $\beta = \nu (d-2+\eta)$\cite{DPS}.

Similar considerations allow one easily to generalize other scaling
laws to account for the new exponent $\Delta$.  For example, the
autocorrelation function $C(\vec x =0, t)$ is readily shown to decay
like $\delta r^{\Delta} t^{-(d-2+\eta)/z}$
as $\delta r \rightarrow 0$.  The exponent $\gamma$ governing
the critical singularity of the equal-time correlation function (i.e.,
the static structure factor), at $\vec k =0$
is given by $\gamma = \nu (2-\eta) - \Delta$.
Note that
our earlier results, and the scaling law relating $\Delta$ to $\beta$,
yield the following results for $\Delta$ at
the mean-field transition and multicritical point, respectively, for
small $D$ and
$d > 2$: $\Delta = 2/(1+\rho)$ and $\Delta = (4+(1+3\rho)\epsilon /2(1+\rho)$.
Of course both these transitions have $\eta =0$ and $\nu =1/2$.

Finally, we verify some of the general results derived above in the
solvable case of $d=0$, $N=1$.  Let us first check the predicted
divergence of the uniform susceptibility over the range $0<r<r_c$,
by solving the $d=0$ problem in the
presence of a uniform field, $h$, where the equation takes the form
$dn/dt = -rn -un^{2+\rho} + h +n \eta (t)$.  As in the case $h=0$,
the Fokker-Planck
equation\cite{VK} for the steady-state
probability distribution function $P(n)$ can be solved
explicitly, with the (Ito representation) result: $P(n) = \frac{1}{Z}
\int_0^{\infty} dnn^{-2r/D-2}
e^{-2h/Dn} e^{-2un^{\rho-1}/D(\rho-1)}$; here $Z$ is a constant chosen
to normalize $\int_0^{\infty} dn P(n) $ to unity.
Given this expression, one
readily calculates $<n>$ as a function of $h$, thereby deriving
a formula for the uniform susceptibility $\chi (h) \equiv
\partial <n> / \partial h$.
In the limit of small $h$ one finds that $\chi (h)$ approaches a finite
limit as $h \rightarrow 0$, provided $|2r/D +1| >1$.  In the band of
$r$ values defined by $|2r/D +1| <1$, however, $\chi (h=0)$ is infinite,
diverging like $h^{|2r/D+1| -1}$ as $h \rightarrow 0$.  Since the
transition from the absorbing to the active phase occurs\cite{Brand,Graham}
at $r_c=-D/2$,
this is consistent with our general argument that $\chi (h=0)$ is infinite
throughout the range $0<r<r_c$ whenever $r_c <0$.  The exact
solution shows that for $d=0$ the divergence actually extends into the active
phase as well.  The explicit dependence on $r$ of
the exponent governing this divergence
as $h \rightarrow 0$ implies that the interval $|2r/D +1| <1$ is controlled
by a fixed line of the RG, with a continuously-varying exponent.
It is easily shown
that $\chi (h)$ diverges logarithmically at the ends of this interval,
and has logarithmic corrections at the critical point $r =r_c$.

Next we use the $d=0$ results to check the scaling theory developed above.
In reference \cite{Graham} one finds that the autocorrelation function
$C(t)$ decays like $\delta r t^{-1/2}$ for $t << \tau$, where the
characteristic time $\tau$ diverges like $\delta r^{-2}$ for small
$\delta r$.  The order parameter decays like $\delta r ^1$.  According to our
earlier definitions, these results
imply exponent values $\Delta =1$, $\beta = 1$,
$\nu z =2$, and $(d-2+\eta)/z =1/2$,
respectively.  It is a trivial matter to verify that these values
satisfy our proposed scaling relation $2 \beta =  \Delta + \nu (d-2+\eta)$.

We gratefully acknowledge conversations with J. Cardy,
C. Jayaprakash,
J. Parrondo.
We are indebted to P. Grassberger
and A. Pikovsky for helpful discussions, and for
sharing the results of unpublished work.


\end{multicols}

\begin{references}
\bibitem{UV} One must introduce an ultra-violet cutoff (e.g., put
the model on a lattice), in order that the integral
$I_d (r)$ be well-defined
at high momenta for
$d \geq 2$.
\bibitem{Ising} For $N=1$ one might also anticipate two phases for the
case, $\rho =1$, which has an apparent Ising symmetry: $n \rightarrow -n$.
If one starts from initial conditions in which $n(\vec x ) $
has the same
sign for all $\vec x$, however, then $n(\vec x ,t)$ will retain that sign
forever after.  One concludes that an active phase with $<n>=0$ but
$<n^2> \neq 0$ is unlikely.
\end{references}
\end{document}